\documentclass[11pt,twoside]{article}
\usepackage{newpasp}

\usepackage{epsf}

\index{replace this with author1's surname, initials}
\index{replace this with author2's surname, initials}
\index{replace this with author3's surname, initials}

\def\Msun{\hbox{$M_\odot$}}
\def\HI{\hbox{H\,{\sc i}}}
\def\HII{\hbox{H\,{\sc ii}}}
\def\CO1_2{\hbox{CO$1\rightarrow 2$}}
\def\H2{\hbox{H$_2$}}

\begin{document}

\title{Cold Molecular Gas, PDRs, \\ and the Origin of \HI\ in Galaxies}
\author{Ronald J.\ Allen}

\affil{Space Telescope Science Institute
3700 San Martin Drive, Baltimore, MD 21218}

\vspace{0.2in}
\hfill \textit{``It's B stars, stupid!''}

% A concise abstract is recommended.  Enter the text of the abstract in
% between the \begin{abstract} and \end{abstract} commands.  Do NOT
% include the word ``Abstract'' in your text; it is insterted
% automatically. Do NOT  make a paragraph break between \begin{abstract} 
% and the first line of the text of the abstract!  Abstracts are required 
% for all papers.

\begin{abstract}
In the currently-accepted model for star formation out of the
interstellar gas in galaxies, the basic construction material is
assumed to be large clouds of atomic hydrogen (\HI). These clouds are
thought to form higher-density complexes of gas and dust, and
turn molecular (\H2). Stars then form out of this molecular gas.

In this paper arguments are advanced for a contrary view, in which the
basic construction material is cold molecular gas out of which the
stars form directly. \HI\ appears in the region when the leftover
\H2\ is illuminated with UV photons from nearby young stars.  The
physics of photodissociation regions provides a natural and
quantitative explanation for the appearance of \HI\ envelopes around
the clouds, and for CO(1-0) emission from the higher-density parts of
their surfaces. In this picture, much of the \HI\ in a galaxy is a
product of the star formation process, not a precursor to it.
\end{abstract}

% Include keywords if you wish. The keywords.apj file, found on aas.org 
% in the pubs/aastex-misc directory, contains a list of keywords used 
% with the ApJ and Letters.  


% That's it for the front matter.  On to the main body of the paper.

% For examples on including figures, see the file vla2000_sample.ps
% at http://www.nrao.edu/vla2000/proceedings/. 
% For examples of figures, equations or tables, please see the file
% vla2000_man.ps at the same site. Also available as
% newpaspman.ps at http://www.aspsky.org/pubs/authors.html

\section{Introduction}

This conference is mainly concerned with the results of observing \HI\
in galaxies, a subject which owes its advancement to the great strides
we have made in radiometers and radio telescopes since the first
detections in the early-1950s. These advances have included the
development of large filled apertures such as the NRAO 300-foot
telescope and the Arecibo Observatory, and especially the great
synthesis-imaging instruments of the world, the Westerbork Synthesis
Radio Telescope and the Very Large Array.

We have been mapping galaxies in \HI\ at ever-increasing resolution and
sensitivity for nearly 50 years, but as far as I know no one has
seriously asked the question:  \textit{Where does the \HI\ come from?}
This question is likely to seem naive whether you are a theorist or an
observer, but probably for diametrically-opposed reasons! If you are an
observer who studies \HI\ in galaxies with the VLA, as most of us here
today are, then you probably assume that the \HI\ is in some sense
\textit{primordial}, i.e.\ it was present in the galaxy before any
stars were formed, and the stars formed from it, and anyone who would
question that point of view must be out of touch with the mainstream.
On the other hand, if you are a theorist who studies the physical state
of the ISM, you know that the time scales for the formation and
destruction of \H2\ in the ISM are so short that essentially all of the
\HI\ in a galaxy has been in the form of \H2\ many times during a
galaxy lifetime, and to ask where the \HI\ comes from in a galaxy is a
naive question similar to ``which comes first, the chicken or the
egg?''

From the point of view of the basic physics, two \HI\ atoms will have a
lower energy state if they can get together and make an \H2\ molecule.
A physicist would therefore expect that, if there is a channel for this
interaction to get rid of its binding energy and angular momentum, and
if there is enough time, in the absence of a source of continual energy
input the gas would all be in the form of \H2. In pure \HI\ gas, there
is a small but non-zero probability that a collision of two \HI\ atoms
will indeed form an \H2\ molecule by emission of quadrupole radiation.
This reaction is favored at low temperatures and high volume
densities.  The re-formation rate can be increased by many orders of
magnitude if dust grains are present in the gas.  \H2\ molecules are
destroyed by UV photons, creating two \HI\ atoms by photodissociation.
The equilibrium state of the gas then depends on the balance between
dissociation and re-formation, and the relative amounts of
N(\H2)/N(\HI) may vary considerably over a galaxy. Indeed, we may
expect to find both \H2\ and \HI\ in various amounts near any and all
sources of far-UV photons.  What observational evidence is there for
this association in galaxies?

Photodissociation regions (PDRs) have been identified in the Galactic
ISM both in the general diffuse gas and in dense regions near hot young
stars.  Theoretical work on PDRs has been stimulated by recent
satellite observations (e.g.\ COBE, ISO, and SWAS), and detailed models
exist especially for the dense surfaces of molecular clouds ($10^2 -
10^7$ cm$^{-3}$) which are illuminated by intense far-UV fluxes ($10^0
- 10^6$ times the local average interstellar radiation field (ISRF)
near the Sun).  Hollenbach \& Tielens (1999) have recently written an
excellent review on this subject with many references. By now we can
say that the basic physics is well understood, and calculations of
infrared line ratios and even line intensities are quite successful at
accounting for the observations under a wide range of physical
conditions in the Galactic ISM.

Perhaps less well known is the fact that the same physics used to
calculate the mid-IR lines of \H2\ from the ISM also provides a
straightforward way to calculate the amount of \HI\ resulting from
dissociation of the \H2\ molecules under the action of the same UV
photons which provide the mid-IR excitation. However, an important
difference between the mid-IR and 21-cm radio observing techniques now
comes into play; in the mid-IR, observational selection favors
high-density, high-UV-flux situations which lead to high surface
brightness. But measurements of the 21-cm \HI\ line favor the
lower-density, lower-UV-flux situations, since in these cases the
\HI\ is spatially more extended and the observational beam filling
factors are therefore larger, making the emission easier to detect with
radio telescopes.

\section{The association of \HI\ with \H2}

\subsection{...in the Galaxy}

Envelopes of \HI\ are often found around Galactic GMCs (1-10 pc scale;
e.g.\ Andersson, Wannier, \& Morris 1991; see Blitz 1993 for a summary)
and have been ascribed to photodissociation of the GMC surface by the
ISRF.  Reach, Koo, \& Heiles (1994) find evidence for \H2\ in some
interstellar cirrus clouds with high column densities (N(\HI) $> 4
\times 10^{20}$ cm$^{-2}$) in quantities about equal to, or greater
than the \HI\, in spite of the fact that these clouds are likely to be
fully exposed to the far-UV flux from the Galactic plane. In the very
few instances where the viewing geometry is favorable, and when the
line-of-sight superposition can be separated using radial velocity
information, \HI\ ``blankets'' can be seen between B Stars \& GMCs
(10-100 pc scale); a clear example of this is Maddalena's Cloud, which
has been described as a large PDR of size $\sim 50 \times 200$ pc
dissociated by far-UV photons from one or two B5-O9 stars each located
about 50 pc from the \HI\ (Williams \& Maddalena 1996).
	
\subsection{...and in other nearby galaxies:}

The first indication that photodissociation may be operating to affect
the large-scale morphology (100-1000 pc) of the \HI\ in galaxies was
found in M83 by Allen, Atherton, \& Tilanus (1986), who noticed a
spatial separation between a particularly well-defined dust lane and
the associated ridge of \HI\ and \HII\ in M83. Other studies have
followed on M83 and on other galaxies (M51, M100; see Smith et
al.\ (2000) for references) and have generally agreed that the initial
interpretation in terms of photodissociation remains a viable option.
The separation in the case of M83 is about 500 pc, and arises because
of the difference between the spiral pattern speed and the rotation
speed of the gas, coupled with the time for collapse of GMCs and the
time that a massive young star lives on the main sequence.

The first study to successfully identify the characteristic PDR
``shell'' morphology of \HI\ in close association with far-UV sources
in a nearby galaxy was carried out on M81 by Allen et al.\ (1997). The
problem is, of course, to obtain sufficient linear resolution ($\sim
100$ pc) in the \HI\ observations to permit one to identify the
morphology of the PDR structures. An important point to note is that
the best ``correlation'' is between the \HI\ and the far-UV,
not between the \HI\ and the H$\alpha$.
 
\begin{figure}[ht]
\plotone{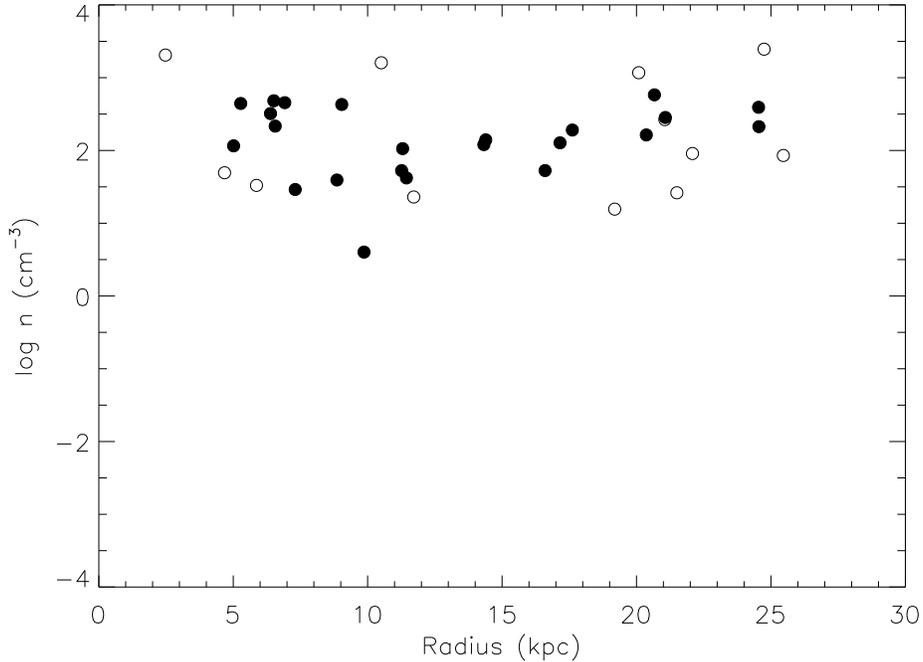}
\caption{\H2\ gas volume density in GMCs near a sample of
35 young star clusters in M101. See Figure 19 in Smith et al.\ (2000) for
further details. \label{fig:m101pdrs}}
\end{figure}

A study similar to M81 but with more quantitative results has recently
been carried out on M101 by Smith et al.\ (2000), who used VLA-\HI\ and
UIT far-UV data to identify and measure PDRs over the whole extent of
the M101 disk. From these observations they derived the volume density
of the \H2\ in the adjacent GMCs in the context of the PDR model.
Figure \ref{fig:m101pdrs} shows the best estimate of the \H2\ volume
densities of GMCs near a sample of 35 young star clusters.  The range
in density (30 - 1000 cm$^{-3}$) is typical for GMCs in our Galaxy,
lending support to the use of the PDR picture, and also shows little
trend with galactocentric distance.

% ++++++++++++++++++++++++++++++++++++++++++++++++++++++++++++++++++++++

% ++++++++++++++++++++++++++++++++++++++++++++++++++++++++++++++++++++++

There is also IR spectral evidence that PDRs are important for
understanding the physics of the ISM in galaxy disks. KAO observations
of the $158\mu$m C\,{\sc ii} line suggest that as much as 70\%-80\% of
the \HI\ in NGC 6946 could be produced by photodissociation (Madden et
al.\ 1993), and ISO spectra in the mid-IR indicate that the bulk of the
mid-IR emission from galaxy disks arises in PDRs (Laurent et al.\ 1999;
Roussel, Sauvage, \& Vigroux 2000; Vigroux et al.\ 1999).

\section{Cold molecular gas in the inner disk of M31}

What evidence do we have that significant quantities of molecular gas
could be ``hiding'' in the ISM of many galaxies? Well, we do know that
many galaxies contain dark matter, and at present there seems to be
nothing wrong with assuming that at least some part of that
dark matter is in the form of a disk of molecular gas. Low-density
molecular gas could escape detection since the excitation of the CO
molecules may be subthermal, but in that case it is not likely that
much mass is involved. If gravitationally-significant amounts of
\H2\ are present, say M(\H2) $\ga$ M(\HI), then the molecular gas would
have to be more dense, and also generally cold in order to escape
detection in the extensive CO(1-0) emission line surveys of galaxy
disks. Such gas could exist in well-shielded parts of galaxies where
the UV flux and cosmic ray density are both very low.

The inner disk of M31 is just such a ``cold'' environment, and it was
here where an initial detection of faint CO emission emanating from
large dark dust clouds was first reported (Allen \& Lequeux 1993).
Modelling of this data (Allen et al.\ 1995) confirmed that the ``cold
gas'' interpretation was viable. Subsequent observations discovered
similar emission in other dust clouds in the inner disk (Loinard,
Allen, \& Lequeux 1996), and also showed that such emission is
ubiquitous in the inner disk of M31 (Loinard, Allen, \& Lequeux 1995),
covering more than half of the annular area between about 1 and 8 kpc
in galactocentric radius. The distinction between cold, extensive gas
and small, warm, beam-diluted condensations was made in favor of the
former with a combination of JCMT CO(3-2) and OVRO CO(1-0)
synthesis-imaging observations on the dust cloud D478 by Loinard \&
Allen (1998). The bulk of the molecular gas in this GMC-sized cloud
appears as an absorbing screen at $\sim 3.5$ K. There is no reason to
presume that D478 is peculiar, so we must conclude that cold molecular
gas is indeed extensively present in the inner disk of M31.

How much mass could be in such a component? Loinard \& Allen (1998)
suggested that if the GMCs were at least marginally bound, then
\H2\ mass surface densities of order $\sim 100$ \Msun\ pc$^{-2}$ would
be present in the inner disk of M31.  However, very recently Pringle,
Allen, \& Lubow (2000) have argued that GMCs are unbound, transient
objects that are not formed from \textit{in situ} cooling of \HI, but
from the agglomeration of the dense phase of the ISM, much of which is
already molecular. How much mass is in this component is presently the
big question, but it's a delicate one that needs some careful thought,
and I am not prepared to address it here any further just yet.

\section{Gas, dust, and young stars in the outer disk of M31}

\begin{figure}[ht]
\plotone{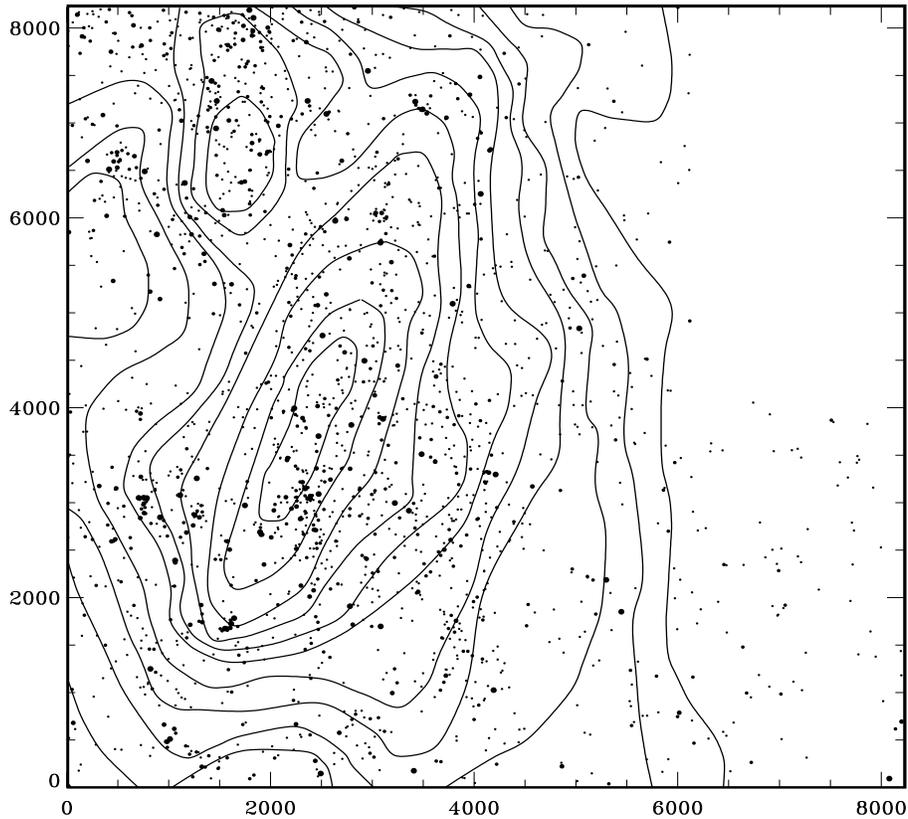}
\caption
{The distribution of blue stars over the extreme outer disk of M\,31
from Cuillandre et al.\ (2000).  This is a $28' \times 28'$ field, and
covers a distance range from 23 to 33 kpc along the SW major axis of
M31. The points represent the locations of stars in the range $20.5 < V
< 24.5$ with $-0.5 < V-I < 0.2$; larger dots are brighter stars.
The contours are \HI\ column density
from Newton \& Emerson (1977), with a resolution of $3.6' \: {\rm EW}
\times 5.8' \: {\rm NS}$ FWHM, and have here been corrected for their
primary beam attenuation.  The \HI\ contours are drawn at levels of 2
through 18 in steps of 2, plus contours at 13 and 19, in units of $ 7.7
\times 10^{19}$ atoms cm$^{-2}$. There is some contamination by quasars
and Galactic white dwarfs, which are visible e.g.\ in the halo field at
the extreme south-west of the image. A small area in the NW corner of this
image shows no stars because this part of the CCD mosaic was not used. \label{fig:M31bluestars}
}
\end{figure}

The presence of significant amounts of molecular gas in the INNER parts
of a galaxy may not be so remarkable in the context of the conventional
wisdom about the state of the ISM in galaxies; however, I think most of
the \HI\ observers would expect that the \HI\ present in the far OUTER
parts of a galaxy is much more ``primordial'', and that star formation
away out there is going to be rare.  Recent results by Cuillandre et
al.\ (2000) contradict both of these bits of conventional wisdom. These
authors used a large-format CCD on the CFHT to observe stars and
background galaxies in a field covering the outer parts of M31 from 23
to 33 kpc, beyond R$_{25}$, a field in which the \HI\ was mapped more
than 25 years ago (Emerson 1974; Newton \& Emerson 1977) with one of
the pioneering \HI\ imaging synthesis radio telescopes, the ``1/2-mile
telescope'' at Cambridge, England. In particular, Cuillandre et
al.\ compared V-I color-magnitude diagrams for \HI-rich regions with
the diagrams for \HI-poor regions, and discovered evidence for dust
mixed in with the \HI\ in amounts corresponding to 0.3 - 0.4 of the
amounts in the Solar neighborhood.  This gas is therefore not
primordial, at least not in the usual sense of not having been
processed in stars. Furthermore, they discovered young B stars
correlated with the \HI, confirming that massive star formation is
going on at the present time. The close association between \HI\ and
the young stars in the outer disk of M31 is shown in Figure
\ref{fig:M31bluestars}. The presence of current, ongoing star formation
implies that \H2\ must therefore also be present, mixed in with the
\HI, in order to form these massive stars.

Could the photodissociation picture developed by Smith et al.\ (2000)
for M101 also work in the outer disk of M31?  A back-of-the-envelope
calculation using the formalism in the Smith et al.\ paper is
encouraging; compact clouds of \H2\ with densities of order 100
cm$^{-3}$ are needed, similar to M101. However, in M31 the field is
clean and unconfused, and a complete census of \textit{all}
far-UV-producing stars can be done. Here we have perhaps the best
laboratory yet for testing the photodissociation picture; a
quantitative analysis is in progress.

% ++++++++++++++++++++++++++++++++++++++++++++++++++++++++++++++++++++++

% ++++++++++++++++++++++++++++++++++++++++++++++++++++++++++++++++++++++

\section{Future perspectives}

I have summarized some of the currently-available evidence in favor of
viewing \HI\ as a product of the star formation process rather than a
precursor to it, and presented a case that \H2\ is likely to be present
in significant quantities in most galaxies. How much \H2\ is really
present in any given galaxy? I believe this is presently an open
question, the answer to which is likely to have important consequences
for several areas of astrophysics.

\acknowledgements

I am grateful to my colleagues at STScI for helping to provide a
scientifically-stimulating atmosphere, and especially to our
theorists Mike Fall and Nino Panagia for discussions on the
ideas described in this paper.

\end{document}